\documentclass[aps,pre,twocolumn]{revtex4-1}
\usepackage{graphicx}
\usepackage{amssymb,amsfonts,amsmath}
\usepackage{color}

\begin{document}

%----------------------------------------------------------------------------------------
%	TITLE AND AUTHORS
%----------------------------------------------------------------------------------------

\title{Weak versus strong wave turbulence in the MMT model} % For titles, only capitalize the first letter

\author{S. Chibbaro$^1$, F. De Lillo$^2$, M. Onorato$^2$}
\affiliation{$^1$Sorbonne Universit\'es, UPMC Univ Paris 06, UMR 7190, Institut Jean Le Rond d'Alembert, F-75005, Paris, France\\
$^2$Dip. di Fisica, Universit\`{a} di Torino and INFN, Sezione di Torino, Via P. Giuria, 1 - Torino, 10125, Italy
}

%----------------------------------------------------------------------------------------

%----------------------------------------------------------------------------------------
%	ABSTRACT, KEYWORDS AND ABBREVIATIONS
%----------------------------------------------------------------------------------------

\begin{abstract}
Within the spirit of fluid turbulence, we consider the one-dimensional Majda-McLaughlin-Tabak (MMT)  model that describes the  interactions of nonlinear dispersive waves.  
We perform a detailed numerical study of the direct energy cascade in the defocusing regime.
In particular, we consider a configuration with large-scale  forcing and small scale dissipation, and 
we introduce three non-dimensional parameters: the ratio between nonlinearity and dispersion, $\epsilon$, and the 
 analogues of the Reynolds number, $Re$, i.e. the ratio between the nonlinear and dissipative time-scales, both at
 large and small scales.
Our numerical experiments show that  
{\it (i)}  in the limit of small $\epsilon$  the spectral slope observed in the statistical steady regime corresponds to
the one predicted by the Weak Wave Turbulence (WWT) theory.
{\it (ii)} As the nonlinearity is increased, the WWT theory breaks down and deviations from its predictions are observed.
{\it (iii)} It is shown that such departures from the WWT theoretical predictions are accompanied by  the phenomenon of intermittency, typical of three dimensional fluid turbulence. We  
 calculate the structure-function
 as well as the probability density function of the wave field  at each scale and show that the degree of intermittency depends on $\epsilon$.
  \end{abstract}

\maketitle % The \maketitle command is necessary to build the title page
%------------------------------------------------

\section{Introduction}
Many physical phenomena are associated with the propagation of dispersive waves. 
While in some cases their dynamics is linear, many relevant situations manifest a non-negligible nonlinearity which produces complex patterns. When the number of degrees of freedom is large enough, such problems must be treated in a statistical manner.
The Weak-Wave Turbulence (WWT) theory  is a very general framework by which the statistical properties of a large number of incoherent and interacting waves can be studied. The  theory, developed during the late sixties 
\cite{zakharov1967energy,falkovich1992kolmogorov,nazarenko2011wave,newell2011wave}, is based on a systematic analytical approach that culminates 
in the so called {\it wave kinetic equation} that describes the evolution of the wave 
spectrum in time (homogeneity and weak nonlinearity are assumed). 
The wave kinetic equation is thus the analogue of the Boltzmann equation for classical particles and, in principle, should be able to give reliable predictions for the statistical distribution of energy as a function of wave numbers, as well as  for various statistical observables.

The WWT theory has been applied to a variety of fields such as for example ocean waves \cite{zakharov1967energy,komen94,falcon2007observation}, capillary waves 
 \cite{pushkarev1996turbulence,falcon2009capillary} Alfv\'en waves  
 \cite{galtier2000weak} or optical waves \cite{picozzi2014optical}.
WWT constitutes hence an interdisciplinary tool suitable for investigating the statistical mechanics of a large number of interacting waves.
A remarkable aspect of such a theory is that, in the presence of an external forcing and dissipation,
exact solutions of the kinetic equation describing constant fluxes of its quadratic conserved quantities can be obtained analytically; 
this important result was achieved for the first time by V. Zakharov in 1966\cite{zakharov1967energy}. 
Despite the beauty of these theoretical results it is of paramount importance to verify if the assumptions behind the theory are realized in practice and thus if this approach is suitable to address physical issues of complex wave systems.

Almost 20 years ago a family of one-dimensional nonlinear dispersive wave
equations, namely the MMT model  \cite{majda1997one}, was introduced as a, in principle simple, model for assessing the validity of WWT theory; 
however, the results reported in \cite{majda1997one} were somehow discouraging and it was reported that ``the predictions of weak turbulence theory fail and yield a much flatter
spectrum compared with the steeper spectrum observed in the numerical statistical steady state''. Deviations from the WWT predictions have been then observed numerically in \cite{zakharov2004one} and have been associated to the presence of coherent structures such as quasi-solitons (see also \cite{rumpf2009turbulent}).
The MMT model has become a ``paradigm-like'' for the verification of the  WWT  predictions.
Indeed, this model offers the rare opportunity to analyse turbulent dynamics over a large range of scales, and it has been proven to display much of the features relevant for turbulent flows \cite{cai1999spectral,cai2001dispersive}, even though it is a one-dimensional idealised model. 

It is instructive to remind that the process of energy cascade observed in
systems of  interacting random waves is similar to the one in fluid turbulence.
Indeed, fully-developed fluid turbulence is known to
exhibit a direct energy cascade with a power-law
energy spectrum very close to $k^{-5/3}$, $k$ being the
spacial wave-number. Such a slope is consistent with the phenomenological Kolmogorov theory  which assumes scale invariance. It turns out that such assumption is not supported by experimental data and numerical simulations; indeed, the three-dimensional fluid turbulence cascade is characterized by  an {\it anomalous scaling} associated to the {\it intermittent} properties of the transfer and dissipation of the kinetic energy \cite{frisch1995turbulence}.
% The two-dimensional fluid turbulence behaves differently; it is characterized by having two cascades (inverse cascade of energy and direct cascade of enstrophy), both of them scale invariant.
Intermittency has been widely studied in fluid turbulence; however, while  exact results have been found for passive scalar dynamics in random flows \cite{fgv}, no analytical prediction (based on the Navier-Stokes equations) for the scaling of the structure functions is at the moment available.

The theory of Weak-Wave Turbulence can be considered as a mean-field approach, where the relevant observable is the wave action spectrum $n(k,t)$, which becomes deterministic in the thermodynamic limit and thus ignores fluctuations.
From this perspective, one would be tempted to rule out the presence of anomalous scaling and intermittency. 
Nonetheless, as anticipated,  many numerical and experimental observations have pointed out 
the existence of such non trivial statistics. 
Notably in wave turbulence, the phenomenon of intermittency has been observed in different experiments of mechanically  forced surface gravity waves by different groups, 
\cite{falcon2007observation,falcon2008fluctuations,lukaschuk2009gravity,nazarenko2010statistics,falcon2010origin}. 
Nonetheless the origin of such intermittency remains mysterious: in 
\cite{lukaschuk2009gravity} such phenomenon has been associated with numerous nonpropagating spikes or splashes and propagating breakers. In experiments reported in \cite{falcon2007observation}  wave breaking or whitecaps  did not occur but cusps (that can leave a signature on the spectrum) were observed on the fluid surface.
In \cite{falcon2010origin} it has been observed that the degree of intermittency depends on the forcing parameter and is somehow independent from breaking and capillary bursts on steep gravity waves.
More in general, in a number of  papers  \cite{New_01a,New_01b,connaughton2003dimensional,Lvo_04,choi2005anomalous,choi2005joint} a tentative 
explanation of the origin of intermittency in wave turbulence has been put forward; however, no quantitative predictions are so far available.

%This discrepancy with theory has motivated the analysis of possible deviations from the premises of the Weak Turbulence, and the presence of intermittency
%\cite{New_01a,New_01b,connaughton2003dimensional,Lvo_04,Cho_05a,Eyi_12}.
%It has also been shown experimentally that theoretical predictions can be largely affected by power injection \cite{falcon2008fluctuations} or energy dissipation 
% \cite{humbert2013wave,miquel2011nonstationary,miquel2013transition}.
%Therefore, the analysis of experiments appears particularly subtle.
%Furthermore, while numerical simulations of realistic set-up are possible\cite{chibbaro2015elastic,deike2014direct,zonta2015growth}, 
%the high resolution requirements for such a kind of simulations raises issues about
%the possibility of a thorough investigation of the foundations of the theory.

In this paper, we consider  the MMT model in the defocusing regime with gravity-wave dispersion as a basic tool for studying the properties of wave turbulence.
While simple, the model has been shown to be dynamically particularly rich, and to display high sensitivity 
to external forces and parameters \cite{cai1999spectral,cai2001dispersive}.
For this reason, here we stick to the same configuration examined in the first  investigation \cite{majda1997one}, to disentangle the various effects. 
In particular, we show that in the direct energy cascade regime the WWT theory offers an accurate prediction for wave spectral slope, provided the ratio between the nonlinear to the linear Hamiltonian is sufficiently small. When the ratio of nonlinear to linear energy is large enough, the prediction of the wave turbulence fails and, as shown in  \cite{majda1997one}, the spectrum becomes steeper than the WWT prediction. The calculation of the structure functions and of the probability density function (PDF) of increments of the wave field at different scales show that, while in the strongly nonlinear regime the dynamics is characterised by intermittency, in the weakly nonlinear case described by the WWT spectrum the wave field exhibits a  quasi-self-similar scaling. The model cannot describe the phenomenon of wave breaking or the formations of cusps; therefore, it offers a unique tool to establish that the observed intermittency cannot be attributed solely to such singular or quasi-singular structures. 
%We believe that present results have broader implications about the intermittent dynamics encountered in interacting waves and provide some explanation for recent results obtained in  numerical experiments \cite{meyrand2015weak,chibbaro2015elastic}.

Our paper is organised as follows: we first describe the MMT model with forcing and dissipation and introduce the control parameters of the simulations. The numerical set-up is then presented and the results on the spectral slopes are shown. We then consider the  structure function and the PDF of the wave field  at different scales and discuss the intermittency properties of the defocusing MMT model. In the last Section, a discussion of the results and the final conclusions are reported.

\section{The theoretical model and the WWT prediction}
We consider the following model
\begin{equation}
i\frac{\partial \psi}{\partial t}= \mathcal{L} \psi+  |\psi|^2\psi + \mathcal{F}+ \mathcal{D},
\label{eq:mmt}
\end{equation}
where $\psi=\psi(x,t)$ and $\mathcal{L}$ is a dispersive operator of the form $\mathcal{L}\exp(i k x)= \omega(k) \exp(i k x)$; the dispersion relation is chosen as  $\omega=\sqrt{|k|}$. $\mathcal{F}$,  $\mathcal{D}$ are two terms that have been included in order to mimic forcing and dissipation; their specific form will be given in the next Section.
The model in (\ref{eq:mmt}) belongs to the  MMT family of equations, which generalizes the nonlinear Schr\"oedinger equation; our selection is based on the fact that, despite it has been shown in \cite{zakharov2004one} that for such member of the family the 
flux for the direct cascade predicted by the WWT has the correct sign and the cascade is local, numerical simulations have shown an unexpected distribution of energy in the spectral modes \cite{majda1997one}.

% (their description will be given later in the numerical details). 
In absence of forcing and dissipation the equation  (\ref{eq:mmt}) preserves  the number of particles $N$,
and the Hamiltonian $H$, 
which can be written as:
\begin{equation}
\begin{split}
%&H=\sum_k \omega_k |\psi_k|^2+\frac{1}{2}  \sum_{k_{1} k_2 k_3 k_4} \psi_1^*\psi_2^*\psi_3\psi_4\delta_{{1+2},{3+4}}
%&N=\sum{ |\psi_k|^2}
H=H_{lin}+H_{nl}=\int \bigg|\frac{\partial \psi^{1/4}}{
\partial x^{1/4}}\bigg|^2dx+
\frac{1}{2} \int |\psi|^4 dx
%\sum_k \omega_k |\psi_k|^2+\frac{1}{2}  \sum_{k_{1} k_2 k_3 k_4} \psi_1^*\psi_2^*\psi_3\psi_4\delta_{{1+2},{3+4}}
\end{split}
\label{eq:H}
\end{equation}
The Hamiltonian is written as the sum of two terms that account for a linear contribution, $H_{lin}$, (first term in {\it r.h.s.})
and a nonlinear one,  $H_{nl}$. 
The crucial assumption of WWT is that  $H_{nl}/H_{lin}\ll1$.
This ratio thus constitutes the small parameter which allows the perturbative approach at the basis
of the development of the WWT theory.
% In the present case, the nonlinear interacting kernel is a homogeneous function of degree 0. 
It is important to underline that in front of the nonlinear term the coefficient is taken as positive: this implies that the model is \emph{defocusing}. 
The case with opposite sign (focusing) is modulationally unstable and its dynamics is dominated by bright solitary waves and coherent structures \cite{cai1999spectral,cai2001dispersive}.
Here, we devote our attention only to the defocusing case, as it has been found to be more pathological with respect to WTT predictions.

By imposing a forcing confined at large scales (small $k$) and a dissipation only at small scales (large $k$), a \emph{direct} cascade in $k$\emph{-space}, characterised by a constant flux of linear energy, $H_{lin}$ is predicted by the WWT once a stationary state is reached. The spectrum is expected to exhibit the following power law: 
\begin{equation}
n_k\sim k^{-1}\;\;\mathrm{WWT\, direct\, cascade}~ ,%C P^{1/3} k^{-1}~,
\label{eq:spectrum}
\end{equation}
 where $n_k=\langle |\psi_k|^2\rangle$ and the brackets $\langle...\rangle$ imply ensemble average.
The verification of such predictions has failed  so far \cite{majda1997one,zakharov2004one,rumpf2009turbulent}.
In particular a spectrum of the following type
\begin{equation}
n_k\sim k^{-5/4}\;\;\mathrm{MMT\, direct\, cascade}~,
\end{equation}
has been revealed in numerical simulations as a new final statistical steady state, consistently with a different closure 
proposed on heuristic grounds. Hereafter, this spectrum will be referred to as the MMT spectrum.
In the following Section 
we will test the validity of the above predictions.

\section{Results}

\subsection{Forcing and Dissipation}
In order to observe the direct cascade, we have performed numerical simulations with the following deterministic instability-type forcing written in Fourier space:
\begin{equation}
\mathcal{F}_k=f \psi_k,\;\; ~k \in [k_{min},k_{max}],
\end{equation}
%and a random forcing, delta-correlated in time.
%\begin{equation}
%\mathcal{F}_k=g(t)\delta(t-t'),\;\; ~k\in [k_{min},k_{max}],
%\end{equation}
where  $f$ is a constant.
% and $g(t)$ is a white noise process.
Dissipation is imposed at very large and small scales using the following terms  
\begin{equation}
\mathcal{D}_k=\mathcal{D}_k^-+\mathcal{D}_k^+=(\nu^-\vert{k\vert}^{-m}+\nu^+ \vert{k\vert}^{n})\psi_k,  
\end{equation}
with $m,n>0$.
%
% in which coherent structures are indeed present but in an intricate way.
%%%%%%%%%%%%%%%%%%%%%%%%%%%%%

\subsection{Control parameters of the simulations}
From equation (\ref{eq:mmt}) it is possible to establish four different time scales
associated with dispersion, nonlinearity, high and low wave number dissipation:
\begin{equation}
\tau_{disp}=\frac{1}{ k_0^{1/2}},\;\; \tau_{nlin}=\frac{1}{| \psi_0|^2},\;\; 
\tau_{diss}^+=\frac{1}{\nu^+ k_0^n},\;\;\tau_{diss}^-=\frac{ k_0^{m}}{\nu^-}
\end{equation}
$k_0$ and $|\psi_0|$ are a characteristic wave number and  amplitude, respectively. 
We can then define the following ``dimensionless'' parameters as the ratio between the dispersive/dissipative  and nonlinear  time scales:
\begin{equation}
%\epsilon_0=\frac{\tau_{disp}}{\tau_{nlin}}=\frac{ |\psi_0|^2}{ k_0^{1/2}},\;\; Re_0^+=\frac{\tau_{diss}^+}{\tau_{nlin}}=\frac{ |\psi_0|^2
%}{\nu^+ k_0^{n}}, Re_0^-=\frac{\tau_{diss}^+}{\tau_{nlin}}=\frac{  k_0^{m}|\psi_0|^2
%}{\nu^-} 
\epsilon_0=\frac{ |\psi_0|^2}{ k_0^{1/2}},\;\; Re_0^+=\frac{ |\psi_0|^2
}{\nu^+ k_0^{n}},\;\;  Re_0^-=\frac{  k_0^{m}|\psi_0|^2
}{\nu^-} 
\end{equation}
where  $Re^{+}$ is the analogue of the Reynolds number in fluid mechanics.
In terms of these parameters, Eq. (\ref{eq:mmt}) takes the following form:
\begin{equation}
i\frac{\partial \psi}{\partial t}=\frac{1}{\epsilon_0} \mathcal{L} \psi+ 
 |\psi|^2\psi + \mathcal{F}+\frac{1}{Re_0^+} \mathcal{D^+}+
 \frac{1}{Re_0^-} \mathcal{D^-},
\label{eq:mmtnd}
\end{equation}
where all the variables and the operators $\mathcal{L}$, $ \mathcal{F}$ and  $\mathcal{D}$ are 
now ``dimensionless''.
Given a computational domain, $\epsilon_0$, $Re_0^{\pm}$ and the forcing amplitude represent 
our control parameters of the simulations. 
In order to extricate the different physical effects,
we shall keep constant the forcing and 
the Reynolds numbers and change only $\epsilon_0$.
The latter parameter indeed controls the ratio between nonlinear to linear waves, and therefore 
represents the key parameter in the WWT theory.
 We  will also monitor the degree of nonlinearity of the asymptotic steady state of 
 our simulation by   considering the ratio of nonlinear to linear Hamiltonian 
 for the equation (\ref{eq:mmtnd}):
 \begin{equation}
 \epsilon=\epsilon_0\frac {\int |\psi|^4 dx}{2\int |\partial \psi^{1/4}/\partial x^{1/4}|^2dx}
 \end{equation}
 Note that the relation between $\epsilon_0$ and $\epsilon$ calculated when a stationary state has reached is not obvious; 
 in the next Section, we will find out that the relation between $\epsilon$ and $\epsilon_0$ is quasi-linear. 

\subsection{Numerical set-up}
Equation (\ref{eq:mmtnd}) has been solved through a Strang splitting pseudo-spectral method. In the 
unforced and undamped case, the method guarantees a conservation of the number of particles and Hamiltonian with a high degree of accuracy. Simulations have been performed
in a periodic box of size $L=2 \pi$ with
a number of
modes  set at $2^{13}$.
The time-marching is carried out with $\Delta t=0.01$.  The coefficients  $\epsilon_0$ and $Re_0^{\pm}$ and $f$ are
selected and the simulation is run until the total  number of particles, the
linear and the nonlinear Hamiltonian reach a stationary state. 
The simulations are performed in the following way: at time $t=0$ the wave-field is initialised with 
$\psi(x,0)=f$. \begin{figure}[ht]
\centering
\includegraphics[width=0.9\linewidth]{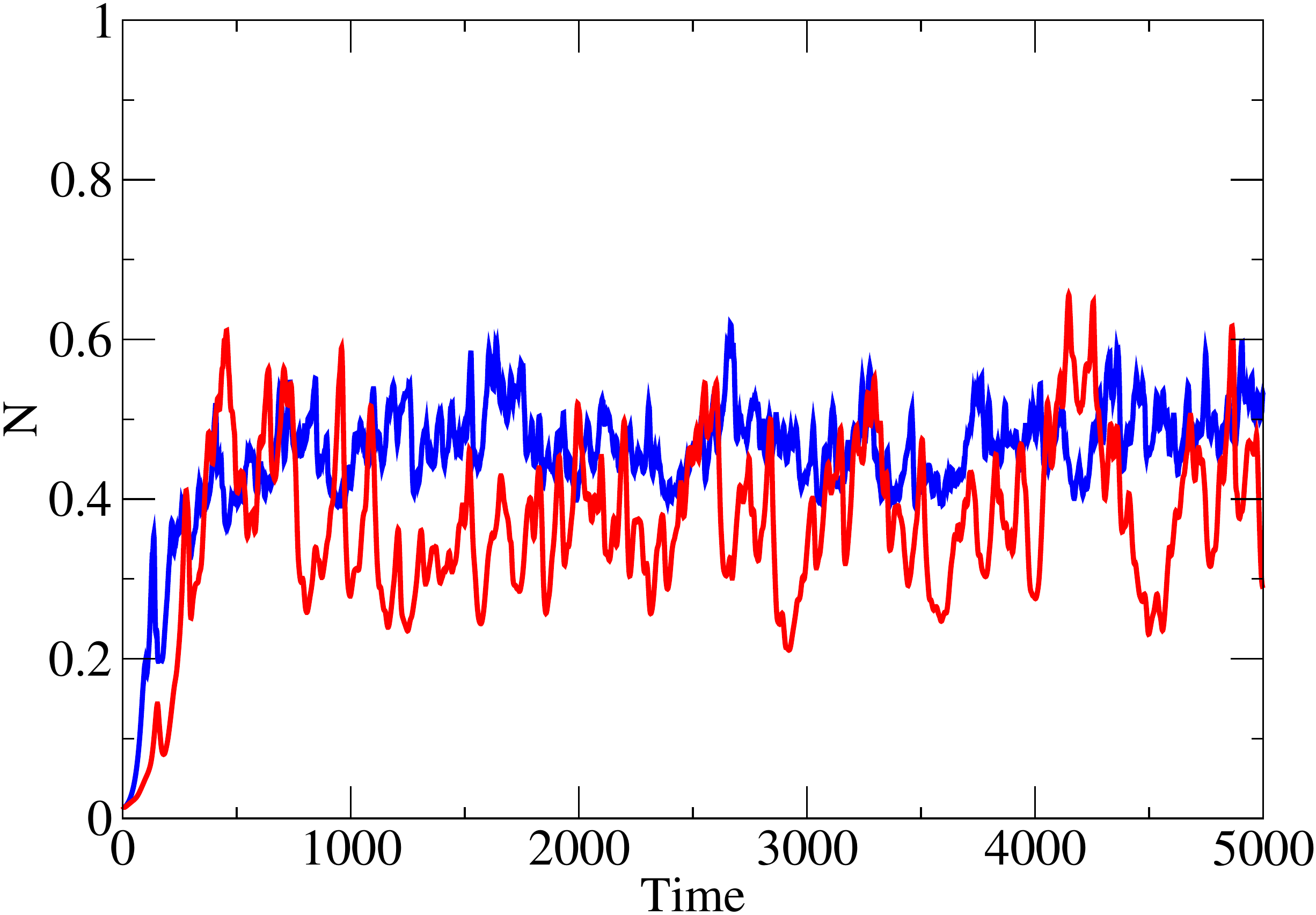}
\caption{Evolution of the number of particles for the simulations 
{\em RUN1} (in blue) and {\em RUN2} (in red).}
\label{fig1}
\end{figure}
Among many of the simulations performed, here, for the sake of clarity, we report two of them, i.e. {\em RUN1} and {\em RUN2}, characterized by different nonlinearity;
the parameters of the simulations are reported in Table \ref{table1}.
In Fig.~\ref{fig1} the evolution of the number of particles 
for two numerical simulations is plotted. The figure shows that in both cases a stationary state is reached.  
It is also possible to appreciate that the fluctuations of the number of particles are not small, especially for the 
more nonlinear case.
%  TABLE
\begin{table}[t]
\caption{Parameters of the two numerical simulations whose spectra are shown 
in Fig.~\ref{fig2}. The values of the power of  the hypo- and 
hyper-viscosity terms are respectively $m=8$ and $n=8$.}
\begin{tabular}{|c|c|c|c|c|c|c|c|c|}
\hline 
  & $f$ & $k_{min}$ & $k_{max}$ & $\epsilon_0$ & $Re_0^+$ & $Re_0^-$   \\
\hline
{\em RUN1} & 0.02 & 4 & 7 & 0.5 & 5$\times 10^{22}$ & 2$\times 10^{-3}$ \\
\hline
{\em RUN2} & 0.02  & 4 & 7 & 12 & 5$\times 10^{22}$ &2$\times 10^{-3}$  \\
\hline
\end{tabular}
 \label{table1}
\end{table}
% END TABLE

After that transient, the
spectra, and all the other statistical observables are computed averaging over time until a satisfactory convergence is reached. 
Over the same
amount of time, the averaged value of $\epsilon$  is calculated in order to
characterize the numerical experiment.
%%%%%%%%%%%%%%%%
\subsection{Spectra}

\begin{figure}[h]
\begin{center}
\includegraphics[width=0.9\linewidth]{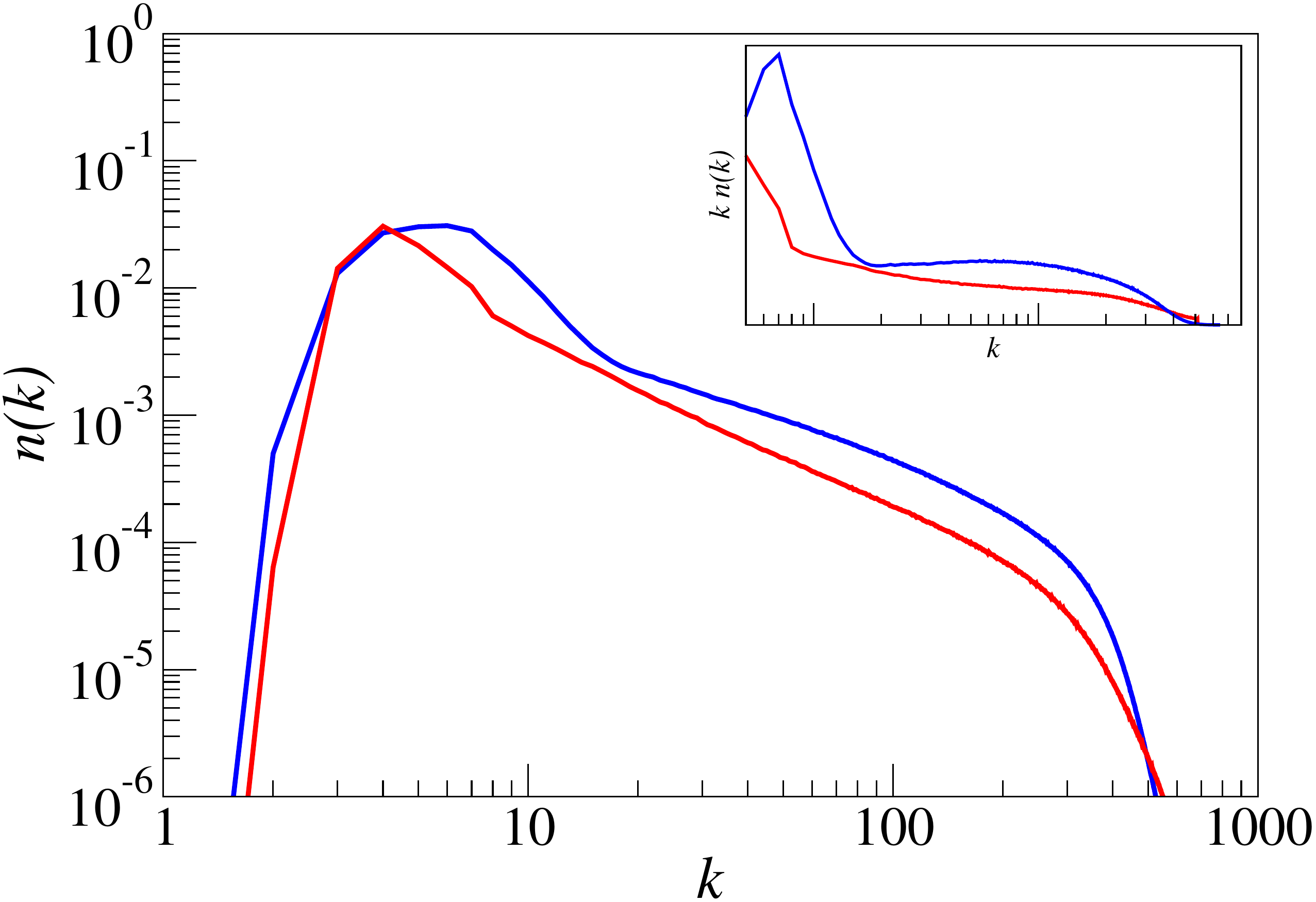}
\caption{Spectra of the two numerical simulations whose parameters are reported in 
Table \ref{table1}. 
In the inset, the spectra are shown after having been multiplied by $k$.
The plot is in linear-log scale. As in all figures, {\em RUN1} is blue and {\em RUN2} red.}
\label{fig2}
\end{center}
\end{figure}

In Fig.~\ref{fig2} spectra obtained averaging for times larger than 10000 are shown.
Both spectra show a power law over an inertial range of  one decade. The figure shows that {\em RUN1}, characterized by a lower value of $\epsilon_0$, displays a spectral slope that is in agreement with the WWT prediction, while the spectral slope for {\em RUN2} appears steeper (see the inset in Fig.~\ref{fig2}). 
Notably, {\em RUN2} evidences a statistical steady state in agreement with the MMT spectrum.

\begin{figure}[h]
\includegraphics[width=0.9\linewidth]{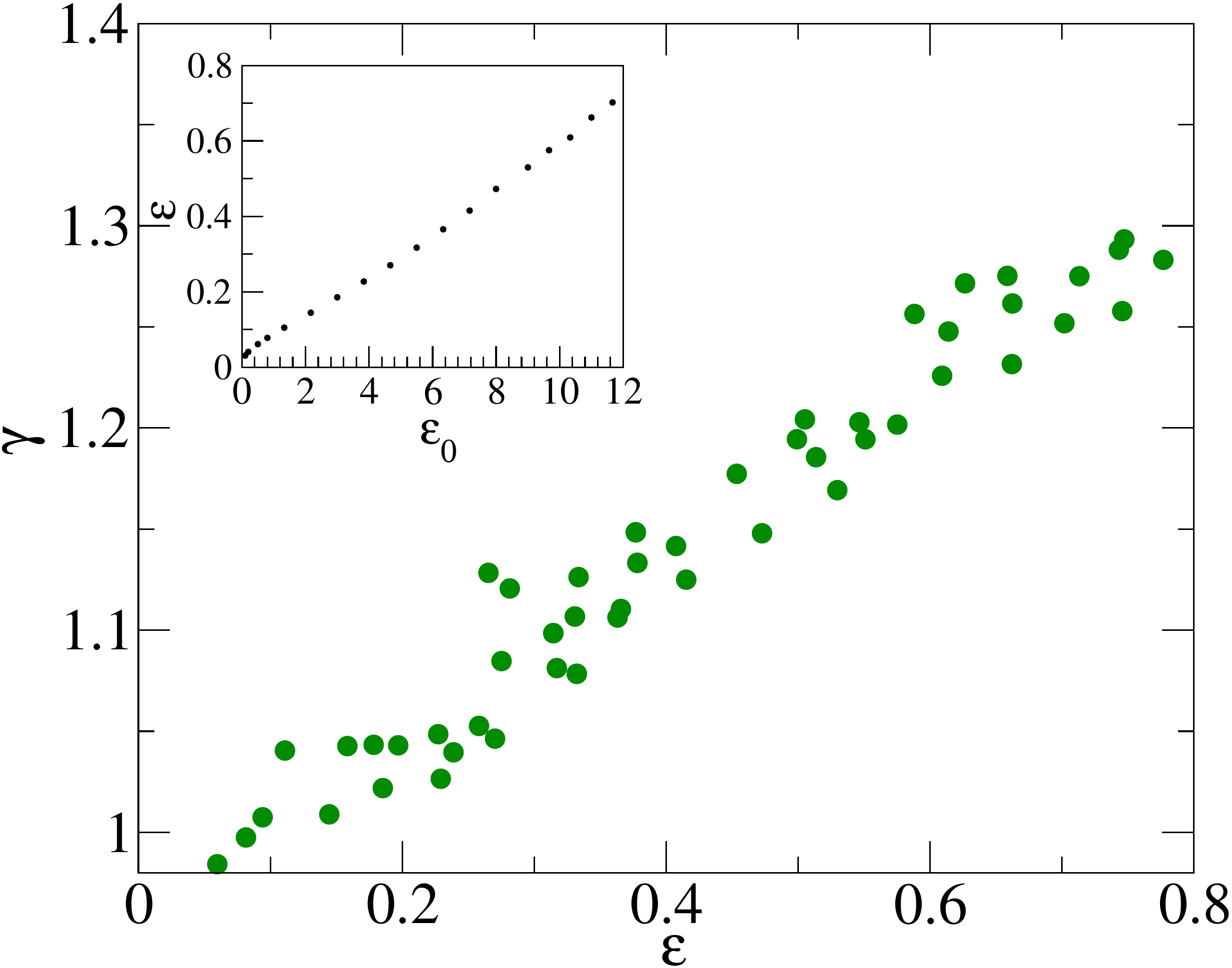}
\caption{Slope of the spectrum as a function of the nonlinear parameter $\epsilon$. In the inset the $\epsilon$ is plotted as a  function of the control parameter $\epsilon_0$.}
\label{fig3}
\end{figure}

\begin{figure}[h]
\begin{center}
\includegraphics[width=0.8\linewidth]{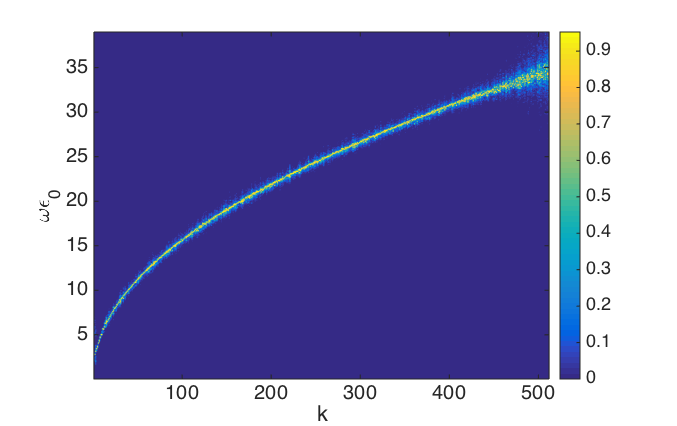}
\includegraphics[width=0.8\linewidth]{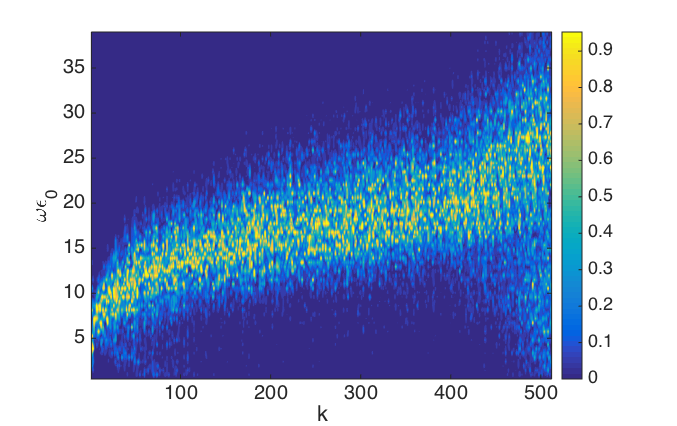}
\caption{ $|\psi(k,\omega)|^2$  (a) for very small nonlinearity $\epsilon\approx0$, {\em RUN1}; (b) for large nonlinearity $\epsilon \simeq 1$, {\em RUN2}.}
\label{fig4}
\end{center}
\end{figure}

In order to appreciate the dependence of the spectral slope, $\gamma$ is the absolute value, on $\epsilon$ we have performed a number of numerical simulations by keeping $f$ and $Re_0^{\pm}$ constant and 
changing $\epsilon_0$ from 0.1 to 12.5; a fit is then performed  in the range of $k\in[20,100]$. 
The results are shown in Fig.~\ref{fig3} where the spectral slope $\gamma$ is plotted as a function of $\epsilon$.
Interestingly, for  $\epsilon \lesssim 0.2$, the slope is in agreement with the prediction of the WWT 
$\gamma=1$. 
Increasing $\epsilon$, the slope of the spectrum changes continuously, attaining the MMT prediction $\gamma=5/4$
for high nonlinearity ($\epsilon \gtrsim 0.6$).
%While a slight saturation of the slope in the range $\epsilon \in [0.6,0.8]$ is found, some simulations display a spectrum even steeper than the MMT.
In the inset
we show the relation between the control parameter $\epsilon_0$ and the effective 
degree of nonlinearity of the simulation
$\epsilon$; for the parameters chosen, the plot indicates 
 a linear dependence between the two numbers. 
%The system reveals a complex dynamics,
%which has a lower bound given by the WWT spectrum,
%and un upper bound given by the critical balance (CB) limit,
%$\tau_{lin}\sim\tau_{nl}$ or $H_{nl}\approx H_{lin}$ and therefore $\epsilon\approx1$~\cite{nazarenko2011wave}:
%equating the linear and nonlinear Hamiltonian parts in (\ref{eq:H}), we obtain
%$\omega_k \psi_k^2  \approx \psi_k^4 k^2 \Rightarrow \psi_k^2 \sim \omega_k k^{-2}$; since we have $\omega_k \sim k^{1/2}$, we obtain the estimate $\psi_k^2 \sim k^{-3/2}$.
%The MMT spectrum could be a better upper bound, but present numerical experiments do not permit to demonstrate this hypothesis.
%  The spectral energy distribution in between the two bounds ($0.2 \lesssim \epsilon < 0.8$
%can be computed only numerically.
%We remark that in each simulation $\epsilon_0$ changes; this leads to different values of  $\epsilon$ and  $Re$. In this respect we have performed extra simulations indicating that if  $Re$ is sufficiently large to guarantee an inertial range, the spectral slope does not depend on $Re$ but only on $\epsilon$.
%
In order to look for other signatures of the transition from weak to strong turbulence we compute the 
quantity $|\psi(k,\omega)|^2$. 
In Fig.~\ref{fig4} we show such quantity normalized in such a way that for each value of $k$ the function 
is scaled so that its maximum is equal to one. It is evident that for {\em RUN1} the linear dispersion relation is nicely followed.
On the other hand, when nonlinearity is high as in {\em RUN2}, the nonlinear interactions lead to a  broadening of the 
frequencies at all scales; in such conditions the WWT theory cannot be applied.

\subsection{Intermittency} \label{sec:int}
Following the studies on the velocity field in Navier-Stokes turbulence \cite{Monin}, we consider as the relevant tool for analysing the intermittent behavior of the wave field the {\it structure functions}, defined as: 
\begin{equation}
S_p(r)=\langle |\delta \psi (r)|^p \rangle =
\int  |\delta \psi (r)|^p {\cal P}(\delta \psi) d\delta \psi
\end{equation}
with $\delta \psi=\Re[\psi(x+r)-\psi(x)]$ where $\Re$ denotes the real part. 
 ${\cal P}(\delta \psi)$ is the PDF of the random field $\delta \psi$. Assuming for example that this probability  is Gaussian at all scales with variance $S_2(r)=\sigma^2(r)$, then we have
 \begin{equation}
 S_p(r)=\sigma(r)^p \int  \left|\frac{\delta \psi (r)}{\sigma}\right|^p {\cal P}\left(\frac{\delta \psi (r)}{\sigma}\right) d\left(\frac{\delta \psi (r)}{\sigma}\right),
 \label{stprob}
 \end{equation}
  where the integral is a constant that depends only on $p$. This property holds for any random
process that is self-similar, i.e. whose statistical properties are scale
invariant. The scaling properties of the structure functions can be related to those of the spectrum by observing that the two-point correlator ${\cal C}(r)=\langle \psi^*(x)\psi(x+r)\rangle$ is the Fourier transform of the spectrum.
For the Wiener-Kintchine theorem~\cite{Monin}, if the wave spectrum is characterized by a
power law $k^{-\gamma}$ with $1<\gamma<3$, then also  $S_2(r)$ scales as $r^{\zeta_2}$, with
$\zeta_2=\gamma-1$. It should be noted that, if WWT is verified, i.e. $\gamma=1$, then  the relation between the exponent of the second-order structure function and the spectral slope does not hold.

 In general, since the spectral  slopes
change for different values of $\epsilon$, we may expect that also  $S_2(r)$ would
scale differently with $\epsilon$. 
Therefore, assuming self-similarity, we can only
make the prediction that the {\it relative} scaling is universal. Indeed,
since $S_p\sim r^{\zeta_p(\epsilon)}\sim S_2^{p/2} \sim r^{
\frac{p}{2}\zeta_2(\epsilon)}$, 
 we expect that  $\zeta_p(\epsilon)/\zeta_2(\epsilon)=p/2$, i.e.
$\epsilon$-independent and linear in $p$.  If the PDF 
is not self-similar, higher moments do not follow any simple scaling relation with respect to the second one and
the integral in (\ref{stprob}) depends on $r$; therefore, we have that
$\zeta_p(\epsilon)/\zeta_2(\epsilon)$ is a nonlinear (possibly $\epsilon$ dependent) function of $p$.
In this case, we encounter an {\it anomalous scaling}, typical of turbulent flows~\cite{frisch1995turbulence}.
From a  technical point of view, the calculation of the scaling exponents 
requires some attention. 
It is rather difficult to get an accurate estimate of
 $\zeta_p$ from the scaling of $S_p(r)$ \emph{versus} $r$. The most appropriate procedure is to plot higher order structure functions versus a reference one and get directly the {\it relative} scaling exponent, hoping in this way to get rid of any spurious dependence due to limited statistics or finite-size effects.
This technique, known as the Extended Self-Similarity (ESS) and developed in the nineties~\cite{benzi1993extended}, has been
widely used in fluid turbulence to analize low-to-moderate Re statistics.
In particular, for three-dimensional turbulence, it is meaningful to plot 
structure functions as a function of $S_3(r)$. The reason is that 
 Kolmogorov $4/5$ law assures, for the Navier-Stokes equations, that $S_3\sim r$. 
 Since we are not aware of any analogous prediction for the present model, we compute the ESS exponents with respect to $S_2(r)$.
%%%%%%%%%%%%%%%%%%%%%%%%%%

\begin{figure}[h]
\begin{center}
\includegraphics[width=0.9\linewidth]{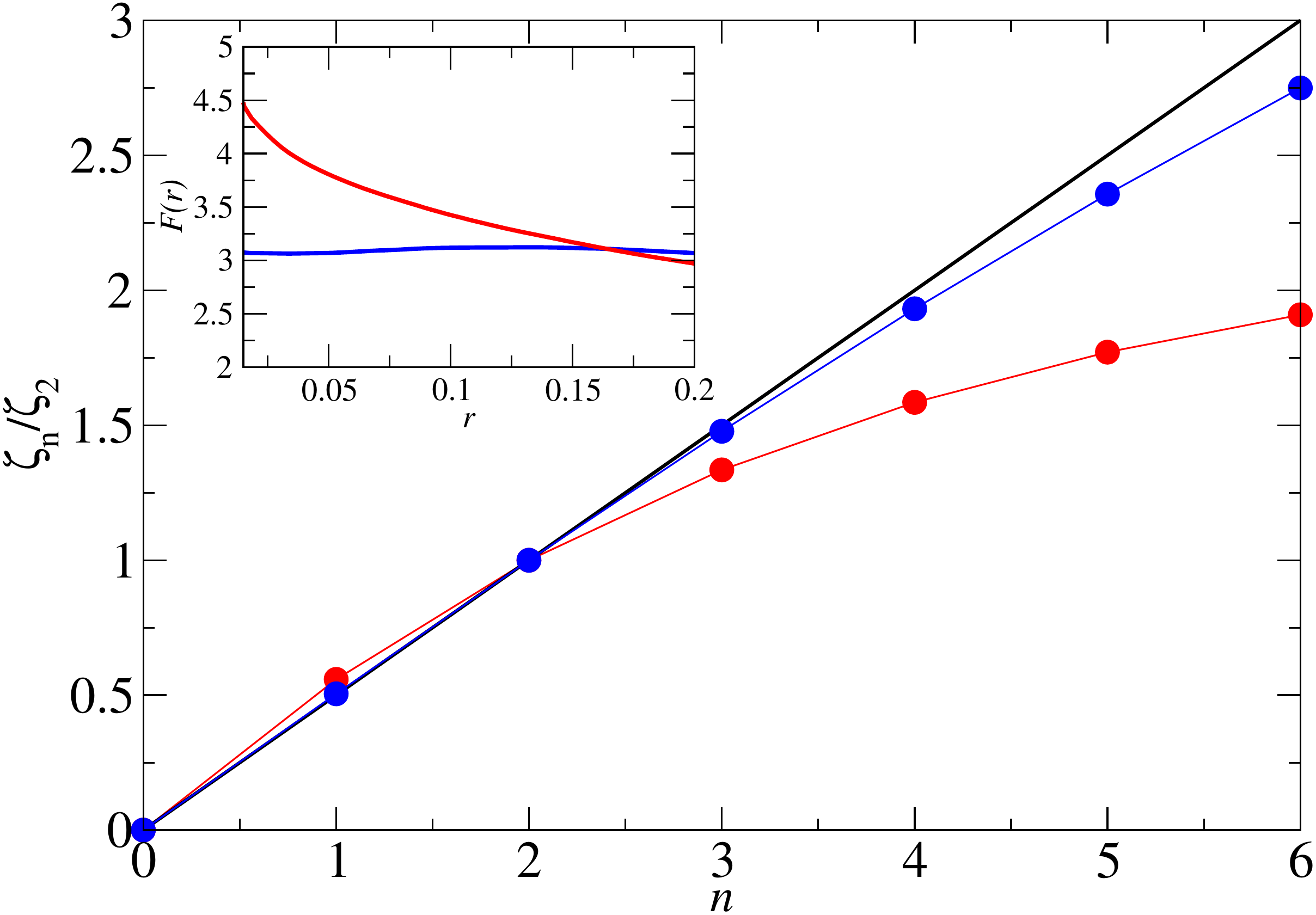}
\caption{The ESS scaling exponents $\zeta_p/\zeta_2$ as a function of $p$ for
{\em RUN1} and {\em RUN2}: while a linear dependence  $\zeta_p/\zeta_2=p/2$ would
indicate self-similarity (consistent with the WWT theory), our simulations show
an increasingly anomalous behaviour as the nonlinearity parameter $\epsilon$ increases. Inset: the flatness $S_4(r)/S_2(r)^2$ (see text). For the weakly non-linear {\em RUN1} the flatness is always close to the Gaussian value 3; higher non-linearities produce more intense fluctuations at smaller scales, the very definition of intermittency.
}
\label{fig5}
\end{center}
\end{figure}

In Fig.~\ref{fig5}, the ratio between  $\zeta_p$ and $\zeta_2$ is shown 
for the two simulations discussed in the previous Section.
We observe that both {\em RUN2} exhibit a degree of intermittency larger than {\em RUN1}. 
The fact that the discrepancy with respect to self similar scaling increases with nonlinearity is not a peculiarity
 of this one-dimensional model under investigation; indeed, our results are consistent with results obtained in very different systems \cite{falcon2007observation,chibbaro2015elastic}.
The nonlinear behaviour of $\zeta_p/\zeta_2$ seen in Fig.~\ref{fig5} implies that
for any structure function of order $p>1$, the ratio $S_{p}(r)/S_2(r)^{p/2}\sim
r^{\zeta_p-p\zeta_2/2}$ increases as $r\rightarrow 0$, or at least down to the
dissipation range. 
In particular, this is true for the flatness
$F=S_{4}(r)/S_2^2(r)$. For a Gaussian field $F=3$, and in general for any
self-similar field F does not depend on $r$. On the other hand, an increasing
flatness at small separation indicates that large fluctuations are relatively
more frequent at those scales, a phenomenon called {\em
intermittency}~\cite{frisch1995turbulence} in the fluid
turbulence literature. It should be noted that $\zeta_p$ is
constrained by general exact results  to be a concave and
non-decreasing function of $p$ for any field with bounded values~\cite{frisch1995turbulence}.
The inset of Fig.~\ref{fig5} shows that the $F(r)$ is constant for the weakly-nonlinear {\em RUN1} (for which $F\approx3$ as in Gaussian case),
while for more non-linear conditions the flatness is observed to increase up to
$\approx 4.5$ at small separations. 
Therefore, the strongly nonlinear case reveals large deviations from self-similarity, with more than $30\%$ of discrepancy in the 6-th order structure function from the WWT monofractal prediction. That is more than what found in Navier-Stokes strong turbulence.
Although the 1-D dynamics may foster such anomalous scaling, present results confirm recent findings in MHD wave turbulence~ \cite{meyrand2015weak}.

\begin{figure}[t]
\begin{center}
\includegraphics[width=0.8\linewidth]{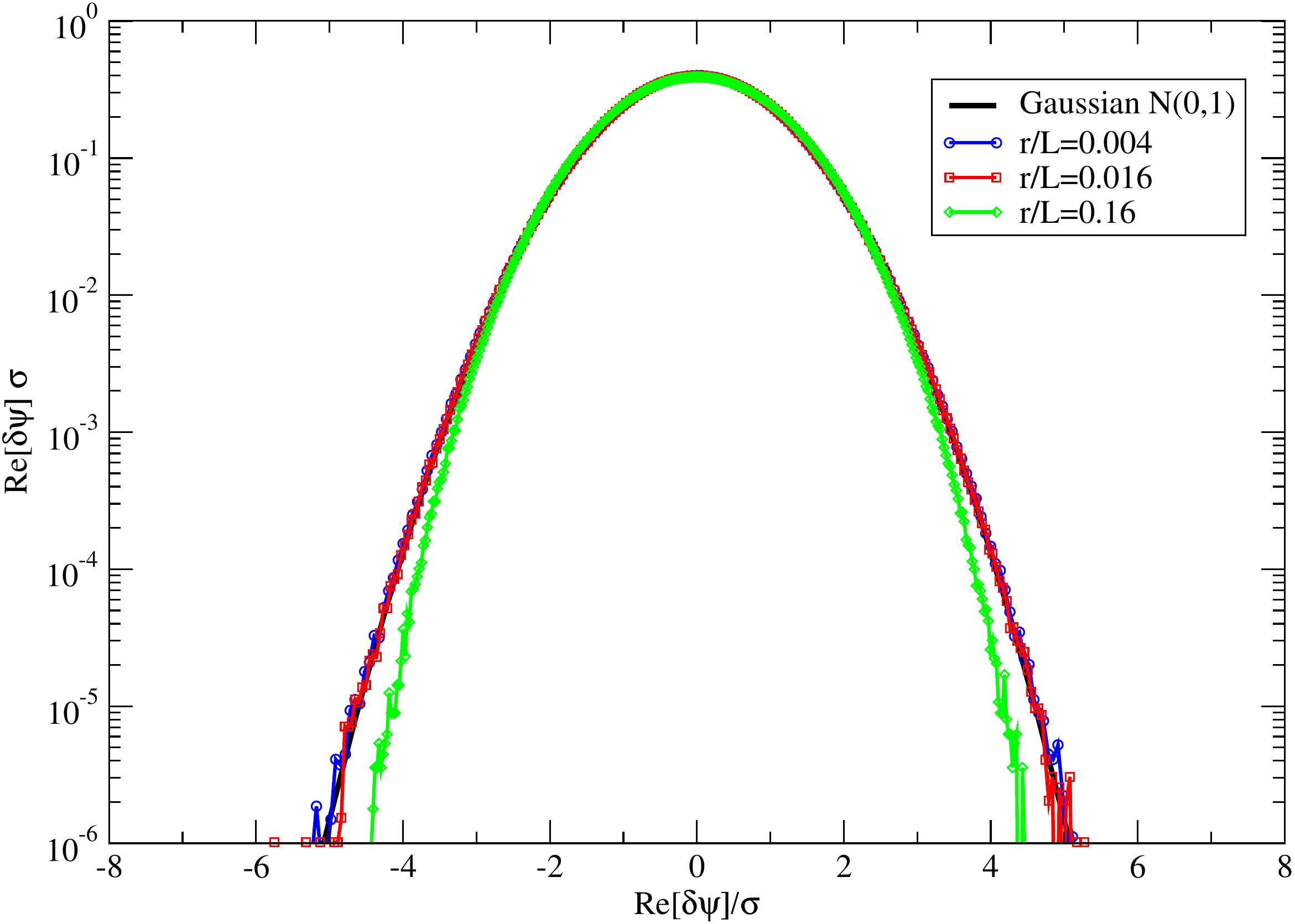}\\
\includegraphics[width=0.8\linewidth]{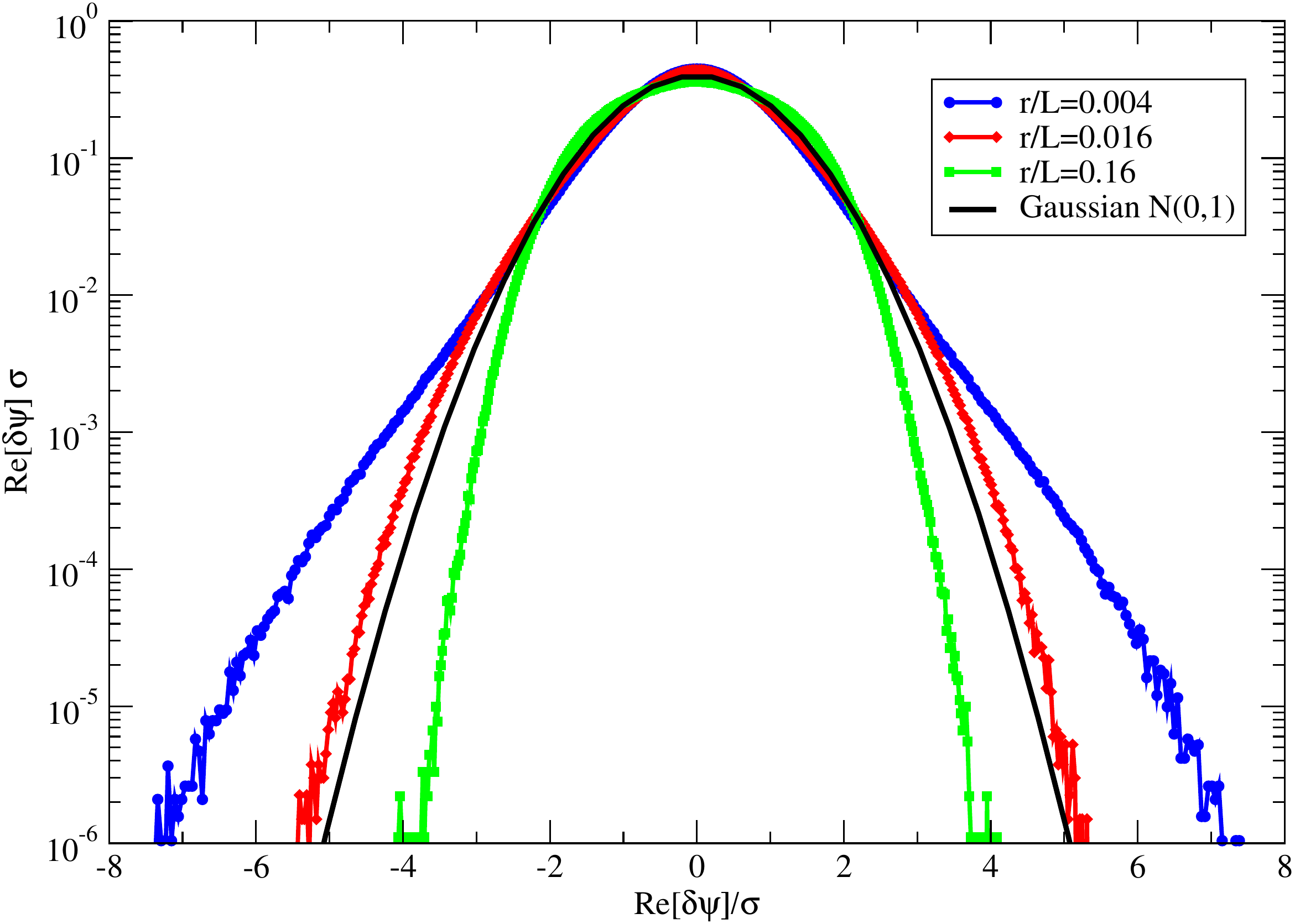}
\caption{PDFs of the increments $\delta_r\psi$ over different values of the separation $r$. Panel(a): {\em RUN1} of Table~\ref{table1}, with a relatively small non-linearity, exhibits a quasi-Gaussian behaviour over the inertial range. Panel (b): {\em RUN2}, with a much larger $\epsilon$, is strongly intermittent, with more intense fluctuations at smaller scales.}
\label{fig6}
\end{center}
\end{figure}

The increasing impact of extreme fluctuations at small scales can be better appreciated by inspecting the PDFs ${\cal P}_r(\delta\psi)$. 
Fig.~\ref{fig6} shows the PDFs of increments for different values of the
nonlinearity $\epsilon$. 
The statistics from {\em RUN1} (low non-linearity) is shown
in panel (a). In this case the PDFs cannot be distinguished  from a Gaussian distribution at all separations but at large ones, where they display sub-Gaussian tails. 
In contrast, panel (b) shows that
{\em RUN2} (high non-linearity) displays  strongly non-Gaussian statistics at small scales, with
wider and wider tails with decreasing separations. 
Gaussian statistics are still found at intermediate separations. Again, 
distributions with sub-Gaussian tails are observed at large scales. 
The scale-dependence of the tails of the distributions is indeed the signature of intermittency. Furthermore this is clearly triggered by  
large enough non-linearity, in agreement with
our previous remarks on Fig.~\ref{fig5}. 
%On the other hand, the sub-Gaussian distributions observed at large separations (corresponding to small wave-numbers) is likely the effect of the pileup of energy on few modes around the forcing scales, where the dynamics is dominated by the injection and is not ``thermalized'' by interactions.

\section{Conclusions}
Our numerical results on the defocusing MMT model with $\sqrt{|k|}$ dispersion indicate that if the nonlinear interactions are sufficiently small, the statistical steady dynamics is well described by the WWT theory. 
Furthermore,  the probability density function of the wave-field is quasi-Gaussian at all scales. 

On the other hand, when nonlinearity starts to become large enough, we have shown that the dynamics is even richer than previously depicted, confirming that the MMT model is a particularly complex  model.
In particular, we have shown that there is a continuous transition from weak to strong turbulence as the  nonlinearity is increased.
In this case, WWT  ceases to represent correctly the system and anomalous scaling is observed, with spectra which differ from the WWT predictions.  
Interestingly, the MMT spectrum is just one of the various possible statistical steady states. 
%Yet, MMT prediction might represent the upper bound for saturated strong turbulence, but our results cannot be said definite. An upper bound is certainly given by the critical balance (CB) estimate. In fact, our results seem to point out that 
%for $\epsilon\approx1$ the spectrum predicted by CB may be attained. Our numerical experiments confirm therefore very recent results  in the transition from weak to strong turbulence at the CB limit  obtained in MHD~\cite{meyrand2016direct}. 

%Concerning the origin of this intermittency, our results seem to indicate 
%that strong nonlinearity generates coherent structures (here solitons) which affects the dynamics at large-scales resulting in a change in spectral energy distribution, but not in a big difference in the probability density of the wave increments. This means that the local structure of the field at large scales is not changed.
%However, this large-scale perturbation  is then amplified through the cascade, which in this sens acts as a multiplicative process, and huge fluctuations are instead present at small scales, with large deviations from gaussianity.
%This picture confirms and give a physical explanation to recent results in elastic turbulence \cite{chibbaro2015elastic}. 

%It will be interesting to study if the nonlinear regime can be included in some way in the Weak Wave Turbulence formalism. 
%Some tentatives, which could work at least for small deviations, are in progress\cite{Lvo_04,Cho_05a,nazarenko2010statistics,Eyi_12} and merit further investigation.

\begin{acknowledgments}
M.O. was supported by MIUR Grant No. PRIN 2012BFNWZ2. 
Dr. B. Giulinico is acknowledged for discussion. D. Proment is acknowledged for discussion 
during the early stages of this work. MO acknowledges Gregory Truden from UNITO who started 
this work as a master student in 2014 at the University of Torino.
\end{acknowledgments}

%\bibliographystyle{unsrt}
%\bibliography{references}

\end{document}